\documentclass[runningheads,a4paper]{llncs}

\usepackage{graphicx}
\usepackage{xspace}
\usepackage{amsmath,amssymb,subfigure}
\usepackage{multirow}
\usepackage{epstopdf}
\usepackage{algorithm}
\usepackage{algpseudocode}

\def \M {\ensuremath{\mathcal{M}}\xspace}

\begin{document}

\mainmatter  

\title{Exact Synthesis of Reversible Logic Circuits using Model Checking}

\titlerunning{Exact Synthesis of Reversible Logic Circuits using Model Checking}

\author{ \emph{Rajarshi Ray} \hspace{0.15in} \emph{Arup Deka} \hspace{0.15in} \emph{Kamalika Datta}}

\authorrunning{Lecture Notes in Computer Science: Authors' Instructions}
\institute{Department of Computer Science and Engineering \\ National Institute of Technology Meghalaya,
Shillong 793003, India\\ Email: rajarshi.ray@nitm.ac.in, arupdekajec@gmail.com, kdatta@nitm.ac.in }

\toctitle{Lecture Notes in Computer Science} \tocauthor{Authors'
Instructions} \maketitle

\begin{abstract}
 Synthesis of reversible logic circuits has gained great attention during the
 last decade. Various synthesis techniques have been proposed, some generate optimal
 solutions (in gate count) and are termed as \emph{exact}, while others are scalable in the sense that 
 they can handle larger functions but generate sub-optimal solutions.
 Although scalable synthesis is very much essential for circuit design, exact synthesis
 is also of great importance as it helps in building design library for the synthesis
 of larger functions. In this paper, we propose an exact synthesis technique for 
 reversible circuits using model checking. We frame the synthesis problem as a model checking instance and 
 propose an iterative bounded model checking calls for an optimal synthesis. Experiments on reversible logic benchmarks
 shows successful synthesis of optimal circuits. We also illustrate optimal synthesis of random functions with as many as 10 variables and up to 10 gates.
\end{abstract}


\vspace{.5cm}


\section{Introduction}

Moore's law \cite{Moore1965} has been witnessed to hold over the past years due to the great advancements in semiconductor fabrication technology, but recently, it is believed to have slowed down. The heat dissipation of the increasing number of transistors in the limited sizes processors, has become a major hurdle in sustaining Moore's law. In 1961, Landauer \cite{Landauer1961} stated that with every bit of information that is lost during a computation, a minimum of $KT \log 2$ Joules of energy is dissipated in the form of heat. In 1973 Bennett \cite{Bennett1973} showed that if zero heat dissipation is required, the computation has to be information loss-less. Reversible computing have gained interest as an alternative to continue Moore's law. Also, since quantum operations are inherently reversible, it can play a vital role in future quantum computing.

A reversible logic circuit consists of a cascade of basic reversible gates, without any fan-out or feedback connections. Also, the number of inputs must be equal to number of outputs, and the circuit must implement a bijective mapping between the inputs and the outputs. Some well-known reversible gates are NOT, CNOT and Toffoli, which constitute the so-called NCT gate library. A reversible gate can be further decomposed into elementary gates known as quantum gates.  The metrics that are generally used to evaluate the quality of a reversible circuit are: (a) The number of gates often called the \emph{Gate Count} (GC), and (b) total number of elementary gates, called the \emph{Quantum Cost} (QC). Given a boolean function to realize, various reversible circuit synthesis methods exist in the literature. These methods can be very broadly classified as either optimal/exact methods \cite{Wille2008,Grosse2009,Wille2012_2,Hung2006}, or sub-optimal methods  \cite{Gupta2006,Datta2012,Datta2012-ISED,Wille2009_1,Fazel2007,Drechsler2011}. Exact synthesis methods are those which generate optimal solutions considering gate count as the metric. In order to guarantee optimality, these methods are generally in-efficient due to the large design space it has to explore, and therefore, can be used for very small functions with very few input lines. Sub-optimal methods on the other hand, are efficient but can provide far from optimal synthesis. For exact synthesis, there are methods that formulate the synthesis problem as a search problem and generates optimal gate netlists \cite{Grosse2009} \cite{Yang2005} \cite{Hung2006}. In \cite{Hung2006}, a solution based on reachability analysis has been proposed. In \cite{Grosse2009}, the synthesis problem has been formulated as a SAT instance and a SAT solver is used to generate an
optimum solution. All these methods are expensive, and provide solutions for small functions. The importance of these methods lie in the fact that they can be used to build libraries of small functions which can be used to construct larger functions. 

In this paper, we propose an exact synthesis method based on a model checking approach \cite{Clarke}. Model checking is an algorithmic technique to verify whether certain properties hold on a system. The system to be verified is modeled mathematically and the property is specified using a specification logic. The contribution in our work is a proposal to model the computations involved in reversible logic synthesis as a Kripke structure \cite{Clarke} and defining LTL/CTL properties such that a falsification of the property on the model provides us with a synthesis, i.e., a sequence of Toffoli gates/permutations. We guarantee an optimal synthesis by searching for falsifying behavior     (sequence of gates) within a fixed length by bounded model checking \cite{biere2003bounded}, and iteratively increasing the bound till we find a falsification for the first time. Apart from being a novel synthesis technique, the advantage of our method is - the limits of functions that can be synthesized exactly can be extended by riding on the advances in model checking techniques. Synthesis of up to $10$ input random circuits can be achieved and results on benchmark circuits are also obtained and compared with other existing optimal methods.

The rest of the paper is organized as follows. Section \ref{back} provides a brief background on reversible logic synthesis and model checking. Section \ref{PW} discuss our proposed synthesis method in detail. In Section \ref{Ex}, we discuss the implementation of our proposed method in the NuSMV model checker and the experimental results. We conclude in Section \ref{con}.


\section{Background}\label{back}

\subsection{Reversible Logic Circuit}

A $n$-input, $n$-output boolean function is called reversible if it implements
a bijective mapping between the inputs and the outputs. In other words, every
input combination maps into a unique output combination, and vice versa.
A non-reversible function can be made reversible through a process called
\emph{reversible embedding}, by adding extra input lines (called \emph{ancilla inputs})
and extra output lines (called \emph{garbage outputs}) as required.
more extra output line known as garbage outputs.
In addition to gate cost and quantum cost, reduction in the number of ancilla and
garbage lines is also often a design objective.

\begin{figure}[htbp]
\centering
\includegraphics[width=3in]{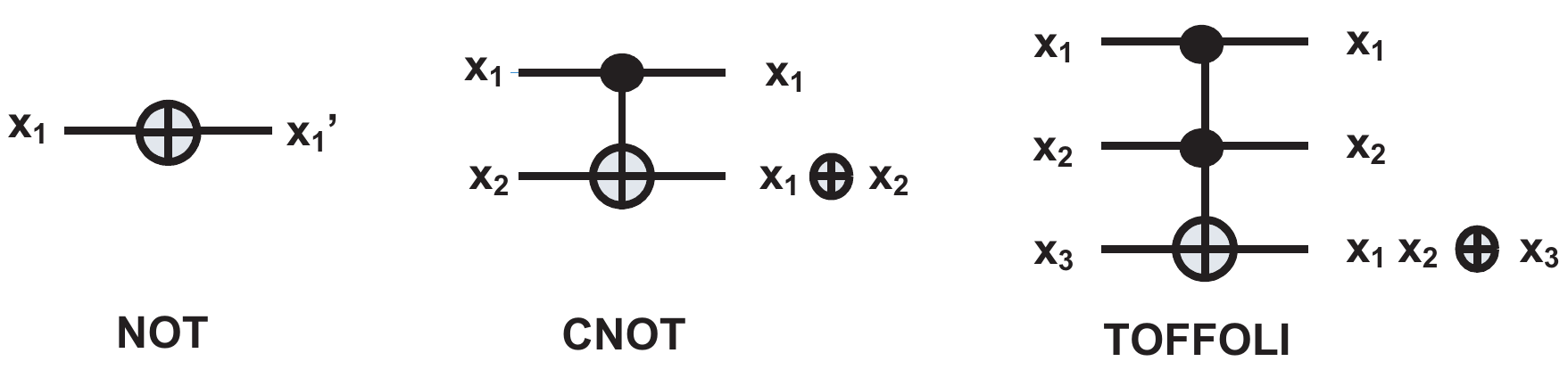}
\caption{Gates of the NCT library}  \label{NCT-gates}
\end{figure}

The bijective mapping implemented by a reversible circuit can also be
represented as a permutation.  For example, the reversible circuit CNOT shown in
Figure \ref{NCT-gates} implements the input to output mapping
as ($00\rightarrow00$, $01\rightarrow01$, $10\rightarrow11$, $11\rightarrow10$).
This mapping can be represented by the permutation P = \{0 1 3 2\}.
In a reversible circuit, the gates are represented by specific
symbols, where $\oplus$ represents the target connection and
a $\bullet$ represents a control connection. The controls are connected to
the target by a vertical line. The logic value on the target line gets inverted
if and only if the logic values on all the control lines are 1.
There are many basic types of reversible gates reported in the literature.
The gates  NOT, CNOT, and Toffoli, as explained below, is a universal set of gates
and constitute the so-called NCT gate library (see Figure \ref{NCT-gates}).
\begin{itemize}
  \item A NOT gate just inverts the particular input and
     produces the output ($x_1 \rightarrow x_1'$).
  \item CNOT or controlled-NOT gate  is a 2-input gate, with control connection
     on line $x_1$ and target line on another line $x_2$. If $x_1$ is 1, the logic value on
     line $x_2$ gets inverted; that is, the gate implements the mapping
     $(x_1,x_2) \rightarrow (x_1,x_1 \oplus x_2)$.
  \item Toffoli gate is a 3-input gate  with two control lines $x_1,x_2$ and one target
     line $x_3$, where the target line is inverted when both the control lines are at 1.
     The gate implements the mapping $(x_1,x_2,x_3) \oplus (x_1,x_2, x_1 x_2 \oplus x_3)$.
\end{itemize}

\begin{figure}[htbp]
\centering
\includegraphics[scale=0.40]{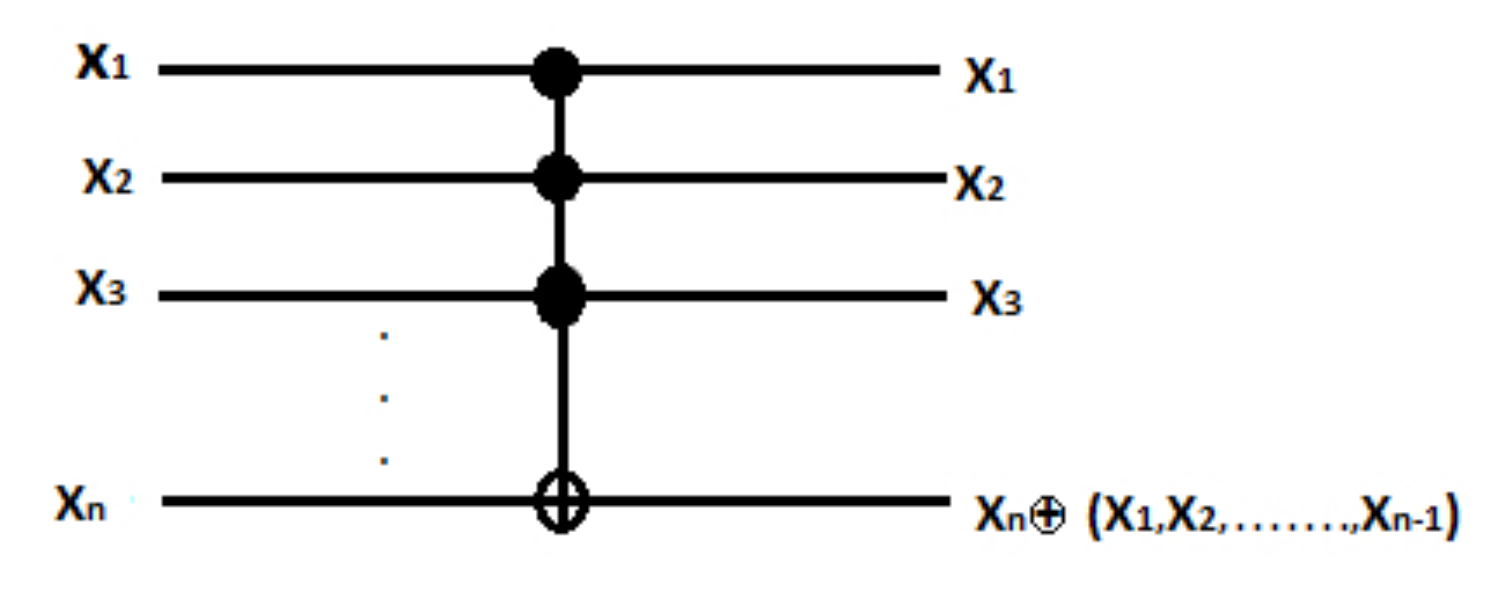}
\caption{A MCT gate}  \label{MCT-gate}
\end{figure}

A \emph{Multiple-Control Toffoli} (MCT) gate is a generalization of the Toffoli gate
with arbitrary number of control lines and one target line, as shown in
Figure \ref{MCT-gate}. The target line gets
inverted when all the control lines are at 1.
In the proposed work, given a reversible function to be synthesized, we first model
the problem formally and then feed the same to a model checking tool. The tool provides
an optimal set of MCT gates to realize the given function as output.

\begin{figure}[htbp]
\centering
\includegraphics[scale=0.50]{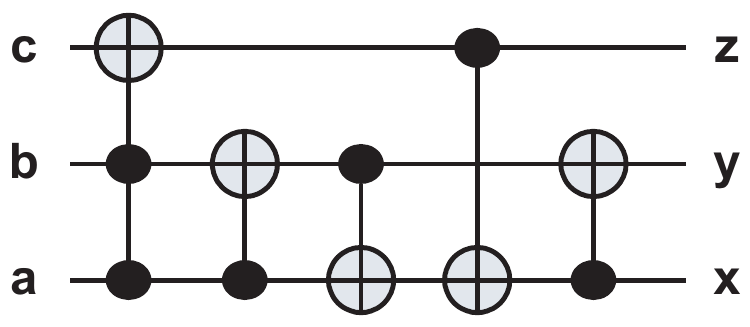}
\caption{An example reversible circuit}  \label{example-gate}
\end{figure}

An example reversible circuit is shown in Figure \ref{example-gate}, which consists of
one Toffoli gate and four CNOT gates.

\subsection{The Optimal Synthesis Problem}
\begin{figure}[htbp]
\begin{center}
\includegraphics[scale=0.40]{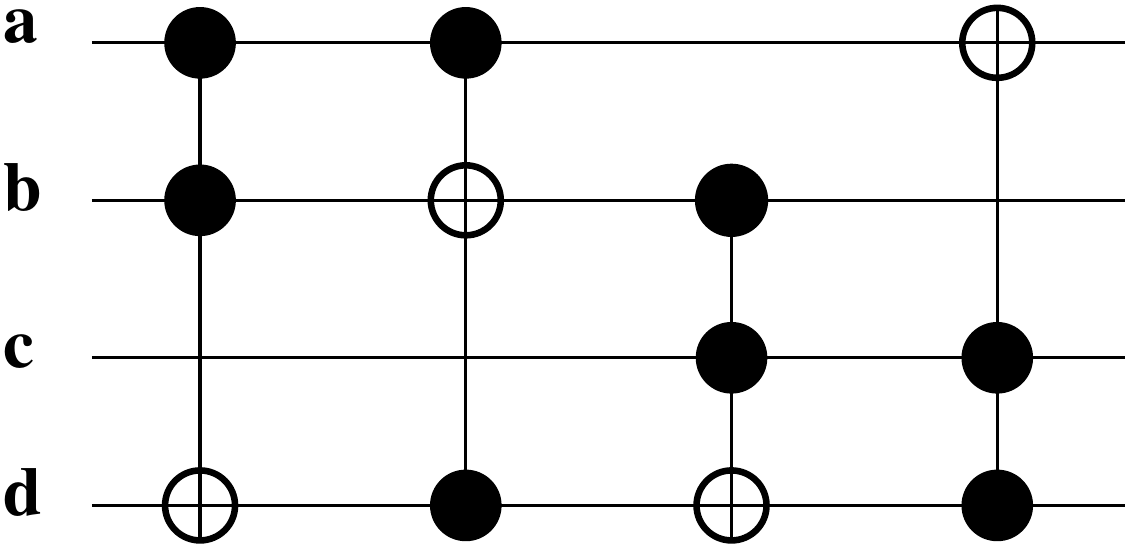}
\end{center}
\caption{Permutation decomposition and synthesis example} \label{decom-ex}
\end{figure}

Suppose that we want to synthesize a 4-input reversible circuit that realizes the permutation
$P = \{ 0,1,2,11,4,5,15,6,8,13,10,14,9,12,3,7 \}$.
The permutation $P$ can be expressed as the composition of four permutations,
$P = P_1 \circ P_2 \circ P_3 \circ P_4$, where
\begin{eqnarray*}
P_1 &=& \{ 0,1,2,3,4,5,6,7,8,9,10,11,13,12,15,14 \} \\
P_2 &=& \{ 0,1,2,3,4,5,6,7,8,13,10,15,12,9,14,11 \} \\
P_3 &=& \{ 0,1,2,3,4,5,7,6,8,9,10,11,12,13,15,14 \} \\
P_4 &=& \{ 0,1,2,11,4,5,6,15,8,9,10,3,12,13,14,7 \}
\end{eqnarray*}
each of which can be realized by a Toffoli gate as shown in Figure \ref{decom-ex}.
Given a final permutation $P$ to be realized as a reversible logic circuit, the synthesis problem that we address is to search for a minimum length sequence of permutations, $P_1, P_2, \ldots, P_n$ such that $P = P_1 \circ P_2 \circ \ldots \circ P_n$. Such a sequence gives as a synthesis of the circuit as a cascade of Tofolli gates $TG_1 \to TG_2 \to \ldots TG_{n-1}$, where each gate $TG_i$ transforms the function $P_1 \circ P_2\ldots P_{i}$ to the new function given by $P_1 \circ P_2 \ldots P_{i+1}$. As we shall explain, it is possible to obtain the Toffoli gate sequence from the permutation sequence and vice-versa.

\subsection{Model Checking}

There are various mathematical formalisms to model a system such as Finite State Machines (FSMs), Kripke models \cite{Clarke}, Binary Decision Diagrams (BDDs) \cite{BLTJ:BLTJ1585} and Ordered Binary Decision Diagrams (OBDDs) \cite{DBLP:journals/tc/Bryant86} to name a few. The commonly used property specification logic used are Linear Time Temporal logic (LTL) \cite{Pnueli:1977:TLP:1382431.1382534} and Computational Tree Logic (CTL) \cite{DBLP:journals/scp/EmersonC82}.We now briefly describe Kripke models, LTL specification logic and the model checking algorithms for LTL.

\subsubsection{A Kripke Model}

\begin{figure}[htbp]
\centering
\includegraphics[scale=0.75]{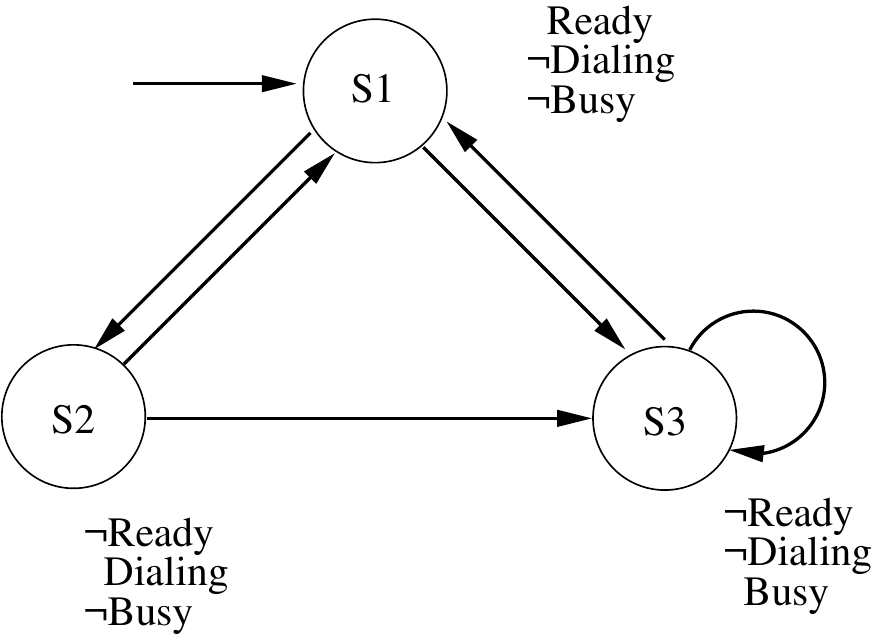}
\caption{Kripke model of a Telephone} \label{mutex}
\label{fig:KripkeModel}
\end{figure}

A Kripke model is a type of finite state machine represented by four tuples $\M (V, S, R, L)$ where $V$ is a finite set
of states, $S \subseteq V$ is the set of initial states, $R \subseteq S \times S$ is a transition relation, and $L$ is a function which
maps every state to a set of atomic proposition which are true in that state. A Kripke model of a simple telephone is shown in Figure \ref{mutex} for illustration. The components of the model $\M (V, S, R, L)$ are as follows:
\begin{eqnarray*}
 V &=& \{S1,S2,S3\} \\
 S &=& \{S1\} \\
 R &=& \{\{S1,S2\},\{S1,S3\},\{S2,S1\},\{S2,S3\},\\&  &\{S3,S1\}, \{S3,S3\}\} \\
 L(S1) &=& \{Ready,\neg Dialing, \neg Busy \}\\
 L(S2) &=& \{\neg Ready, Dialing, \neg Busy \} \\
 L(S3) &=& \{\neg Ready,\neg Dialing, Busy \} \\
\end{eqnarray*}

The telephone has three states $S1$, $S2$, $S3$ and three atomic propositions
\emph{Ready}, \emph{Dialing} and \emph{Busy}. Initially, the telephone is in the state $S1$ having only the proposition $Ready$ as $true$. From $S1$, it can either take a transition to $S2$ having $Dialing$ as true or it can take a transition to $S3$ having $busy$ as true by receiving an incoming call. From the state $S2$, the telephone may either take a transition to state $S3$ when there is a connection or it may take a transition to $S1$ in case connection cannot be established. From $S3$, there can be either a transition to itself or a transition to $S1$ when the call ends.

A property to be verified on the model of a telephone is that it should not be in any state where both $Dialing$ and $Busy$ are true.

\subsubsection{Linear Time Temporal Logic}

Temporal logic provides a specification of the ordering of different events in the system without explicitly specifying the time \cite{Pnueli:1977:TLP:1382431.1382534}. LTL properties are evaluated over computation paths which are infinite sequence of states of the model. An example of a path starting from state $S_1$ in the model of Figure \ref{fig:KripkeModel} is $\pi = S_1 \rightarrow S_2 \rightarrow S_3 \rightarrow S_3 \rightarrow \ldots$. An LTL formula has the following syntax \cite{Michael-Huth}:
\begin{equation}
 \phi := \top \mid \bot \mid \psi \mid (X\phi) \mid (F\phi) \mid (G\phi) \mid (\phi_1 U \phi_2)
\end{equation}
where $\psi$ is any propositional formula and $X$, $F$, $G$ and $U$ are the basic temporal connectors. For a path $\pi = s_1 \rightarrow s_2 \rightarrow s_3 \rightarrow \ldots$, $\pi^{i}$ shall represent the suffix of the path starting from the state $s_i$ in $\pi$. The satisfaction relation $\models$ between $\pi$ and an LTL formula $\phi$ is defined below \cite{Michael-Huth}:

\begin{itemize}
\item $\pi \models \textbf{X}\phi$ \textit{ iff } $\pi^{2} \models \phi$. 
\item $\pi \models \textbf{G}\phi$ \textit{ iff } $\forall i \geq 1$, $\pi^{i} \models \phi$.
\item $\pi \models \textbf{F}\phi$ \textit{ iff } $\exists i \geq 1$ \textit{ such that } $\pi^{i}\models \phi$.
\item $\pi \models \phi_1 \textbf{U} \phi_2$ \textit{ iff } $\exists i\geq1$ \textit{ such that } $\pi^{i} \models \phi$  \textit{ and } $\forall j = 1,\ldots,i-1$, $\pi^{j}\models \phi_1$.
\end{itemize}

A model $\M$ satisfies $\phi$ \textit{iff} all computation paths of $\M$ starting from an initial state satisfies $\phi$. The LTL model checking algorithm consists of constructing a Kripke model $\M_{\neg \phi}$, ($\neg$ is the logical not operator), with Buchi accepting condition, i.e., a path in the model is accepting \textit{iff} it has some final state repeating infinitely often. 
The construction of $\M_{\neg\phi}$ is such that it accepts precisely those paths which satisfies $\neg\phi$. This model is then combined with $\M$ resulting in a model whose paths are present both in $\M$ and $\M_{\neg\phi}$. If there is no such combined path, we deduce that $\M$ satisfies $\phi$. The presence of a combined path implies a computation path in $\M$ that does not satisfy $\phi$ and is thus a counter-example to the specification.


\subsubsection{State Space Explosion}
The construction of the Kripke model $\M_{\phi}$ for an LTL formula $\phi$ is $O(|\phi| \cdot 2^{|\phi|})$. Computing the product of $\M_{\phi}$ and $\M$ and then checking for a path in the product results in the complexity of $O(|V+E| \cdot 2^{\phi})$. Therefore, LTL model checking algorithm is exponential in the size of the formula and linear in the size of the model. The labeling algorithm for CTL model checking performs state labeling starting from the simplest sub-formulas iteratively. If there are $k$ connectives in the 
formula, the model has to be labeled $k$ times. The labeling of formula 
$AG\phi$ and $EG\phi$ are expensive as it requires performing a breadth first search 
for every vertex of the model. This gives the complexity of the algorithm as 
$O (k\cdot V \cdot (V + E))$, $V$ and $E$ being the number of vertices and 
edges in the model respectively. The complexity can be reduced to 
$O (k \cdot (V + E))$ by computing the strongly connected components followed 
by breadth first search for labeling the expensive connectors as mentioned earlier. 
Although CTL model checking is linear in both the size of the formula and the 
model, the size of the model itself grows exponentially with the number of variables 
of the system. This exponential growth of state-space is known as the state space 
explosion problem. Various methods have been proposed to overcome this problem and 
implemented in the model checkers 
\cite{Burch:1992:SMC:162045.162046,EClarkeWNZ}. 

We see that the size of our proposed model of reversible logic synthesis computationis exponential in the number of input variables of the function to be realized.

\section{Proposed Synthesis Method}
\label{PW}
A reversible circuit with $n$ inputs can be represented as a pair of input-output permutation of elements in the integer set $\{ 0, 1, \ldots , 2^{n}-1\}$. The input permutation is usually fixed to $(0, 1, \ldots, 2^{n}-1)$ showing the decimal representation of all possible inputs in an $n$ input circuit. The output permutation is a permutation of the elements in the set $(0, 1, \ldots, 2^{n}-1)$  showing the decimal representation of the output lines corresponding to the inputs. The optimal synthesis problem is to find a minimum sequence of MCT gates which realizes the output permutation from the input permutation, if feasible. In this work, we attempt the synthesis problem by proposing a Kripke model such that a path in the model shows the evolution of permutations from the input permutation corresponding to the application of a sequence of MCT gates. The goal then is to find a computation path in the model from the initial state to a goal state corresponding to the desired output permutation. The existence of such a path is searched by a model checker when directed to do so with a LTL/CTL specification formula.

In our proposed model, there are two components which are composed synchronously. One of the components model the selection of a MCT gate out of all the possible gates in an $n$ input logic. The other component models the transition from the current permutation to the next permutation when the selected gate in the other module is applied. The composition models the computation of permutations. We now present our proposed models for the mentioned two components.

\subsection{Model of MCT Gate Selection} \label{GE}

\begin{figure}[htbp]
\centering
\includegraphics[scale=.50]{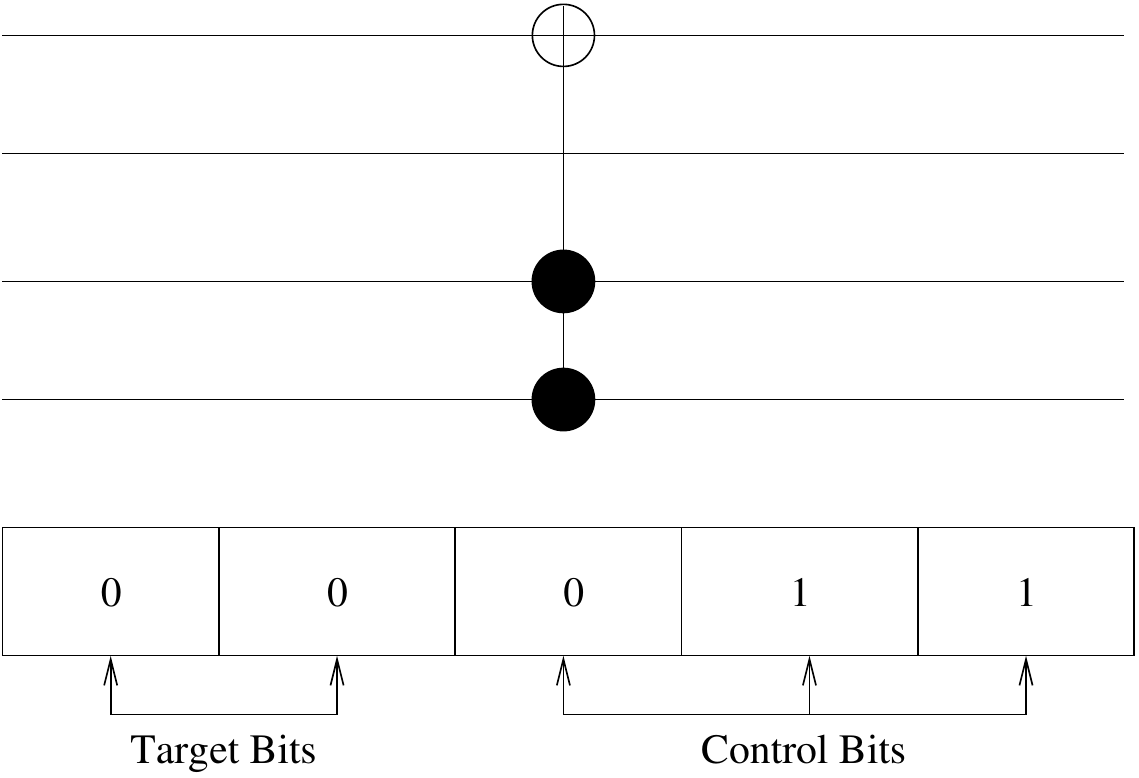}
\caption{Encoding of a 4-input MCT gate}
\label{fig:GateEncode}
\end{figure}

There are $n.2^{n-1}$ MCT gates possible with an $n$ input logic. We use a 
boolean encoding to uniquely encode each of the MCT gates. Our encoding 
uses $\lceil \log_2 n \rceil$ bits to encode the position of the target line in 
the gate since it can be in any of the $n$ lines. In the remaining $n-1$ lines, a 
control may or may not be present which is encoded with a bit per line. In total, 
we use $\lceil \log_2 n \rceil + n - 1$ bits to encode a MCT gate. The 
$\lceil \log_2 n \rceil$ most significant bits (msb) are used to encode the target 
line number and the remaining bits encode the control lines of the MCT gate.
Figure \ref{fig:GateEncode} shows an example of a 4 input MCT gate. Encoding of 
the target line position needs $\lceil \log_2 4 \rceil = 2$ bits. Our 
encoding counts the first input line of the MCT from above as line number 0 
and encodes the target line there by 00 in the two msb bits. For representing the 
controls, the encoding use the remaining 3 bits. The absence of control in line 1 
and the presence of control in line 2, line 3 is encoded by 011.

\begin{figure}[htbp]
\centering
\includegraphics[scale=.50]{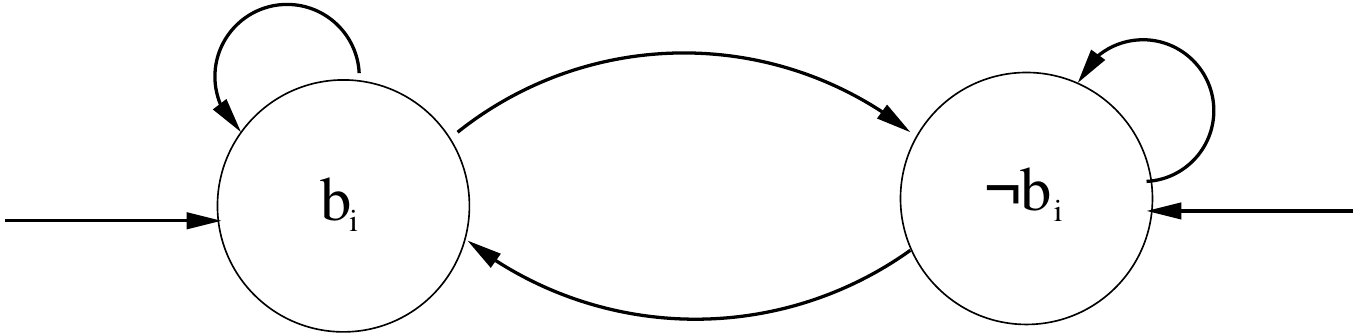}
\caption{Kripe Model of a bit value selection}
\label{fig:randomBit}
\end{figure}

Using the above encoding, a MCT gate selection can be modeled by selecting bit 
entries of $b_0,b_1,\ldots,b_{\lceil \log_2 n \rceil + n - 2}$ at random. We 
model this by having a Kripke model per bit $b_i$ as shown in Figure 
\ref{fig:randomBit}. The model chooses to either flip or retain $b_i$ 
non-deterministically. The model of the gate selection is then obtained by 
composing the models for $b_0$ to $b_{\lceil \log_2 n \rceil + n - 2}$. 
Note that the size of the composed model is $O(2^{n})$ where $n$ is the number of 
bits in the MCT Gate encoding. 

\subsection{Model of Permutation Transition}
\label{MPT}
\begin{figure}[htbp]
\centering
\includegraphics[scale=.75]{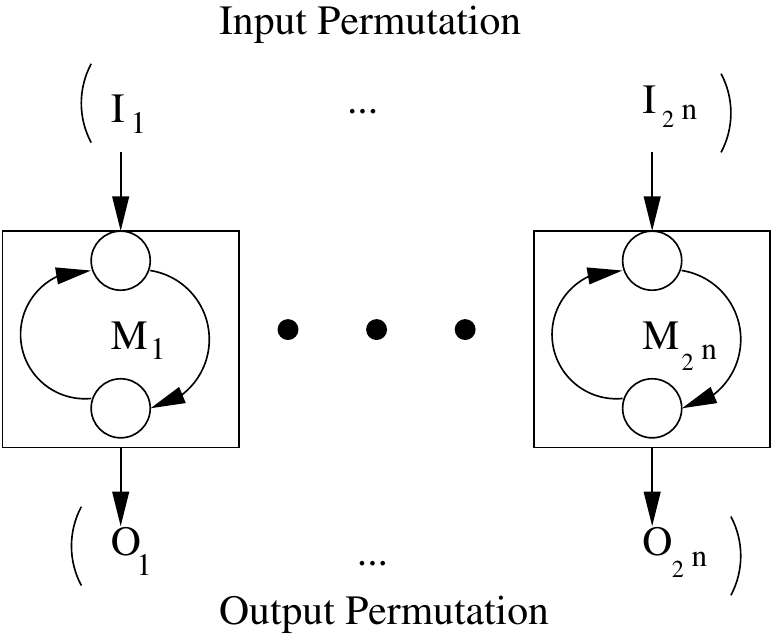}
\caption{Transition using Kripke models}
\label{fig:transition}
\end{figure}

Given a valuation of the $n$ input lines and an encoding of a selected MCT gate to be applied to the input lines, the transition of the input  to the next valuation is computed using a Kripke model. Every input to output transition ($2^n$) is computed individually by independent transition models. All the models are composed as depicted in Figure \ref{fig:transition} to compute the transition of the input permutation. 

The atomic propositions of the transition Kripke model are the inputs 
$i_0,\ldots,$ $i_{n-1}$, and the encoding of the selected MCT gate given by 
$t_1,\ldots,t_k,c_0,\ldots,c_{n-2}$. In total, there are 
$\lceil \log_2 n \rceil + 2n - 1$ propositions in the model. The initial 
state of the transition model has the initial values of the input lines as
$i_0,\ldots,i_{n-1}$. Observe that the 
application of any possible MCT gate may flip the values of only the input 
proposition $i_{tg}$ where $tg$ is the target line of the gate keeping the 
remaining propositions fixed. The transitions of the model is given by the 
\emph{next} relation shown in Eqn. (\ref{eqn:next}) assuming that 
$tg \in \{0,\ldots,n-1\}$ is the target line specified by 
variables $t_1,\ldots,t_k$ of the gate encoding.

\begin{equation}\label{eqn:next}
next(i_k) =
\begin{cases}
  i_k & tg \neq k \\
	i_k \oplus \bigwedge_{j=0}^{n-2}(\neg c_j + i_\ell) & \textit{otherwise}. \\
\end{cases}
\end{equation}
where $\ell = j$ for $0 \leq j < tg$ and $\ell = j+1$ otherwise. $+$  and 
$\bigwedge$ denote boolean OR and AND operation respectively. It can be 
easily verified that the \emph{next} transition relation models the application 
of the selected MCT gate on the input. In this way, the above model in composition 
with the gate selection model emulates the computation of states beginning with 
the state with the initial permutation to the next possible permutations.

\subsection{Specification in LTL and CTL}

In order to carry out logic synthesis using the model checker, we feed it with the model 
of the computation along with a specification in LTL or CTL formula. The 
property we specify is \emph{there does not exist any computation path from the 
given initial permutation to the goal permutation state}. If the model checker finds 
the specification to be false, then it produces a counter-example which is precisely 
the sequence of MCT gate-ids to be applied to the input permutation in order to 
obtain the required output permutation. On the other hand, if the model checker finds the 
specification true then it implies that no synthesis is possible for the given 
input-output permutation pair. Let $\phi$ be the propositional formula which is 
true only in the state where the output permutation is true (the values of 
$i_0,\ldots,i_{n-1}$ in the output permutation), then we use the future operation in 
LTL and specify $\neg F (\phi)$ to specify  that a path does not have any future 
state where the goal permutation is true. When this LTL formula is given to a 
model checker, it checks the property for all the computation paths beginning from 
the initial state of the model. Similarly, we specify the CTL formula 
$\neg EF(\phi)$ to specify the negation of there exist a computation path from 
the initial permutation state which has a future state with the goal permutation 
as true.

\subsection{Optimal Synthesis}

It may be noted that the
reversible logic synthesis method using above mentioned LTL/CTL 
specification gives a solution which may not be optimal. For 
optimal synthesis, we use bounded model checking, whereby we verify the 
specification on paths of bounded length. We increment the bound iteratively,
and look for counter-example from the model checker at every step. 
The smallest bound for which the model checker finds a counter-example 
is guaranteed to produce a logic synthesis solution using  minimum number of gates. 
The procedure is outlined in Algorithm \ref{algo:optimalSynthesis}. The function 
\emph{MC.get()} in line 8 is used to retrieve a counter-example from the model checker.

\begin{algorithm}
\caption{Optimal Logic Synthesis with Bounded Model Checking}
\begin{algorithmic}[1] 
\Procedure{Optimal Synthesis}{MC, \M,	$\phi$}\Comment{$\phi$ is the LTL/CTL spec, MC is a Model Checker}
	\State $bound \gets 0$ 
	\For {$bound \leq MAXBOUND$}
			\State $c \gets $MC(\M,$\phi$)
			\If{$c$ is true}
			  \State bound++
			\Else
				\State $Counterexample \gets MC.get(\M,\phi)$ \label{alg:getC}
			\EndIf
	\EndFor
\EndProcedure
\end{algorithmic}
\label{algo:optimalSynthesis}
\end{algorithm}
	
\section{Experimental Results}\label{Ex}

We implement our model as described in the previous section, in the input 
language of the NuSMV model checker \cite{DBLP:journals/sttt/CimattiCGR00}. 
NuSMV is an open source model checker with a large user community. The input language allows modular 
implementation to ease modeling of complex systems. Every NuSMV file must 
have a \emph{main} module. Different components of the system can be 
implemented as separate \emph{modules}, similar to functions in C 
language. We implement the gate selection and the transition component as 
separate modules described in the earlier section. An LTL formula in NuSMV is specified using the \emph{LTLSPEC} keyword. 

The implementation takes the number of variables
and the desired goal permutation as the inputs, and
generates the optimal gate sequence as output. In the implementation, we have considered $(0, 1, 2, 3, \ldots, 2^n-1)$ as 
the initial permutation, where $n$ denotes the number of inputs. 
Given the goal permutation, we generate the LTL specification that there 
exist no sequence of MCT gates that can realize the goal permutation. The model checker NuSMV then checks for the correctness
of the claim and returns a sequence of MCT gates, if the assumed claim is found to be false. If the assumed claim is true, we know
that the synthesis is not possible.

Since our proposed method is an exact synthesis method, we
have compared the synthesis results with other reported exact synthesis methods
\cite{Grosse2009,Datta2012,DmitriMaslov} only. The results are summarized in Tables \ref{res-inp-3} for input size of
3, and Table \ref{res-inp-456} for input sizes of 4, 5 and 6. 
The experiments are performed on 
an Intel core-i3 desktop with 2.40 GHz and 4GB main memory, running Ubuntu v14.04.
The tables only show the benchmarks reported in the previously published papers are shown.
In the tables, the first column shows the name of the benchmark where known,
while the second column shows the number of inputs $n$. 
The next three columns show the gate count as reported in \cite{Datta2012},
\cite{Grosse2009} and \cite{DmitriMaslov} respectively, where the entries
marked as `$-$' indicate that values have not been reported.
The last three columns of the tables show the gate count (GC), quantum
cost (QC), and run time using the proposed algorithm. We observe that the minimum gate count
synthesis obtained with our proposed solution matches with the previously reported methods.
This establishes the correctness of our method. As evident from the results, we could also synthesize 
a number of benchmarks optimally, for which previous works did not report any solution. It is not known
whether they were not reported because the earlier methods could not obtain solution on these benchmarks or these were skipped 
in their evaluation.  

Table \ref{res-random} shows synthesis results for many randomly generated
6, 7, 8, 9 and 10 size input permutations. Note that the permutations are  
chosen such that the optimal gate count to synthesize these is bounded by 10.  
The experiments illustrate that our method could generate minimum gate
solutions for input size of 10 variables, when the optimal solution is bounded by 10 gates, within reasonable memory and time
constraint.

\begin{table}
\caption{\small{Result of the experiments input size = 3}} \label{res-inp-3}
\centering
\begin{tabular}{|l|l||c|c|c|c|c|c|}
\hline
\multirow{2}{*}{Name}& \multirow{2}{*}{\begin{tabular}[c]{@{}l@{}}Given\\
permutation\end{tabular}} &
\multicolumn{3}{c|}{\begin{tabular}[c]{@{}c@{}}Gate Count\\
(GC)\end{tabular}}             & \multicolumn{3}{l|}{Proposed
Algorithm}   \\ \cline{3-8}
                                                                         &    & \multicolumn{1}{c|}{\cite{Datta2012}} & \multicolumn{1}{c|}{\cite{Grosse2009}} & \multicolumn{1}{c|}{\cite{DmitriMaslov}} & GC & \multicolumn{1}{c|}{QC} & Time (sec)  \\ \hline
Peres & 0,3,2,5,4,7,6,1 & -- & 2  &-- & 2  &  6  & 0.04\\ \hline
Fredkin & 0,1,2,5,4,3,6,7 & -- &  3 & --& 3  &  7  & 0.06\\ \hline
Ham3 & 0,7,4,3,2,5,1,6 & 5  &  5 &-- & 5  &  9  & 0.09\\ \hline
N$^{th}$prime & 0,2,3,5,7,1,4,6 &-- & --&-- & 4 & 8 & 0.07\\ \hline
Ex1 & 4,5,6,1,0,7,2,3 &-- &4& --& 4 & 16   & 0.06\\  \hline
&   1,0,3,2,5,7,4 6 &4 &-- &-- & 4  & 8  & 0.07\\ \hline
&   7,0,1,2,3,4,5,6 &3 &-- &-- & 3  & 7  & 0.05\\ \hline
Miller & 0,1,2,4,3,5,6,7 & 5&5 &-- & 5  & 9 & 0.08\\ \hline
&   1,2,3,4,5,6,7,0 &3 &-- &-- & 3  & 7  & 0.05\\  \hline
&   3,6,2,5,7,1,0,4 &7 &-- & --& 7  & 19 & 0.18\\  \hline
&   1,2,7,5,6,3,0,4 &6 &-- &-- & 6  & 14 & 0.12\\  \hline
&   7,5,2,4,6,1,0,3 & 7&-- &-- & 7  & 19 & 0.17\\  \hline
&   7,6,5,4,3,2,1,0 &-- &-- &-- & 3  &  3 & 0.05\\  \hline
&   4,3,0,2,7,5,6,1 &6 &-- &-- & 6  & 10 & 0.12\\  \hline
3-17 & 7,1,4,3,0,2,6,5 & 6&6 &-- & 6  & 14 & 0.09 \\ \hline
\end{tabular}
\end{table}

\begin{table}
  \caption{Synthesis results for $n$ = 4, 5 and 6}  \label{res-inp-456}

  \begin{center}
    \begin{tabular}{|l|c||c|c|c||c|c|c|} \hline
      Name & $n$ & \multicolumn{3}{c|}{GC using} & \multicolumn{3}{c|}{Proposed approach} \\ \cline{3-8}
           &     & \cite{Datta2012} & \cite{Grosse2009} & \cite{DmitriMaslov} & GC & QC & Time (sec) \\ \hline \hline
      1-bit-adder &   4  &  5  & -- & -- &  4 & 12 & 0.11 \\ \hline
      2-to-4-decoder& 4  & --  &  6 & -- &  6 & 18 & 0.30 \\ \hline
      decode-42   &   4  & --  & -- & 10 & 10 & 30 & 45.83 \\ \hline
      Mperk       &   4  & --  & -- &  9 &  9 & 17 & 2.04 \\ \hline
      Imark       &   4  & --  & -- &  7 &  7 & 19 & 0.49 \\ \hline
      4\_49       &   4  & --  & -- & 12 & 12 & 72 & 12659.30 \\ \hline
      hwb4        &   4  & --  & -- & 11 & 11 & 23 & 232.88 \\ \hline
      Oc5         &   4  & --  & -- & 11 & 11 & 39 & 704.10 \\ \hline
      Oc6         &   4  & --  & -- & 12 & 12 & 44 & 14408.50 \\ \hline
      4gt5        &   5  & --  &  4 & -- &  3 & 19 & 0.33 \\ \hline
      mod5\_mills &   5  & --  &  5 & -- &  5 & 13 & 0.70 \\ \hline
      mod5d\_1    &   5  & --  &  7 & -- &  7 & 15 & 6.84 \\ \hline
      mod5d\_2    &   5  & --  &  8 & -- &  8 & 16 & 5.77 \\ \hline
      mod5        &   5  & --  & -- & -- &  7 & 15 & 6.23 \\ \hline
      4mod5\_younus&  5  & --  & -- & -- &  5 &  9 & 0.71 \\ \hline
      4mod5\_miller&  5  &  5  & -- & -- &  5 & 13 & 0.78 \\ \hline
      graycode    &   6  & --  &  5 & -- &  5 &  5 & 2.22 \\ \hline
      permanent2x2&   6  & --  & -- & -- &  3 & 23 & 1.24 \\ \hline
    \end{tabular}
  \end{center}
\end{table}

\section{Conclusion}\label{con}

In this paper, an exact synthesis method has been proposed for reversible logic
circuits. The synthesis problem has been framed as an instance of model checking
and the NuSMV model checker is used to generate the solutions. Experimental results
on benchmarks verifies that the method indeed generates optimal circuits (a cascade 
of MCT Gates of minimum length). Many random functions having as many as 10 input 
10 variables could be optimally generated, having up to 10 MCT gates. This shows that 
our method does scale to handle even large functions. Though, results to synthesize
benchmarks of large size (8 or more input variables) remains to be verified.

\begin{table}
  \caption{Synthesis results on random permutations with 6, 7, 8, 9 and 10 inputs} \label{res-random}

  \begin{center}
    \begin{tabular}{|l||c|c|c|} \hline
      Name & \multicolumn{3}{c|}{Proposed approach} \\ \cline{2-4}
           &  GC  &  QC  &  Time (sec) \\ \hline \hline
	   $random\_1$(n=6)&    4  &   48   &  0.669           \\ \hline
	 $random\_2$(n=6)&    5  &   25  &  0.877           \\ \hline
	$random\_3$(n=6)&    6  &   78   &  1.340           \\ \hline
	 $random\_4$(n=6)&    8  &   84  &  9.253           \\ \hline
	 $random\_5$(n=6)&    10  &   122  &  76.809           \\ \hline
	$random\_1$(n=7)&    4  &   52   &  2.440           \\ \hline
	$random\_2$(n=7)&    5  &   75   &  3.182           \\ \hline
	$random\_3$(n=7)&    6  &   63   & 4.635           \\ \hline
           $random\_4$(n=7)&   7  &  179   &  7.169          \\ \hline
	$random\_4$(n=7)&   8  &  192   &  11.224          \\ \hline
	$random\_4$(n=7)&   9  &  188   &  140.065          \\ \hline
           $random\_5$(n=7)&   10  &  174    &  279.483           \\ \hline
	$random\_1$(n=8)&   4   &  143    &  10.199          \\ \hline
          $random\_2$(n=8)&   5   &   184  &  12.791           \\ \hline
          $random\_3$(n=8)&   6  &  114    &  17.091          \\ \hline
          $random\_4$(n=8)&   8  &  160   &  47.674           \\ \hline
	 $random\_5$(n=8)&   10  &  213    &    2333.793       \\ \hline
	 $random\_1$(n=9)&   6  &  79   &    84.316       \\ \hline
	 $random\_2$(n=9)&   8  &  110   &    174.604       \\ \hline
	 $random\_3$(n=9)&   10  &  373   &    3679.524       \\ \hline
	 $random\_1$(n=10)&   6  &  175   &    414.778       \\ \hline
	 $random\_2$(n=10)&   8  &  215   &    1018.712       \\ \hline
	 $random\_3$(n=10)&   10  &  501   &    6110.31       \\ \hline
    \end{tabular}
  \end{center}

\end{table}

\bibliographystyle{unsrt}

\end{document}